\documentclass{article}

\usepackage{graphicx}

\usepackage{hyperref}
\usepackage{amsmath}
\usepackage{amssymb}	
\usepackage{mathtools}

\usepackage{subcaption}
\usepackage[dvipsnames]{xcolor}

\usepackage{multirow}
\usepackage{pdfpages}
\usepackage{stfloats}
\usepackage{tabularx}
\usepackage[export]{adjustbox}
\usepackage{float}
\usepackage[
singlelinecheck=false 
]{caption}

\begin{document}
\title{\bf Effect of quintessence dark energy on the shadow of Hayward black holes with spherical accretion}
\author{{Malihe Heydari-Fard \thanks{Electronic address: heydarifard@qom.ac.ir} }\\{\small \emph{ Department of Physics, The University of Qom, 3716146611, Qom, Iran}}}

\maketitle

\begin{abstract}
It is expected that the astrophysical black holes are surrounded by a luminous accretion flow that is a necessary ingredient for imaging a black hole. In this paper, we study the influence of quintessence dark energy on the shadow images of a Hayward black hole surrounded by the static/infalling spherical accretion flow. We find the effect of the state parameter of quintessence matter on the horizons, photon sphere and impact parameter of the quintessence Hayward black hole. The observed specific intensity of the shadow and also the shadow and photon ring luminosities of the quintessence Hayward black hole in two different spherically accretion flow are investigated. We also use the Event Horizon Telescope observational data of Sgr A* and M87* to constrain the free parameters of quintessence Hayward black hole. Finally, by comparison the results of quintessence Hayward black holes with quintessence Schwarzschild and Hayward black holes we find that the effect of quintessence matter on the black hole shadow is more significant than the regularity effect.
\vspace{5mm}
\end{abstract}

\section{Introduction}
\label{1-introduction}
The Event Horizon Telescope (EHT) collaboration has released Very Long Baseline Interferometry (VLBI) observations at the core of the M87 galaxy \cite{A1}--\cite{A6} and Milky Way galaxy \cite{A7} with angular resolution comparable to that expected of a supermassive black hole. The EHT images show a dark central area called the black hole shadow which is surrounded by a bright ring, the photon ring \cite{Cunha1}. These results consistent with predictions of the theory of general relativity (GR) \cite{Book} and thus the EHT observations provide another strong evidence to Einstein’s theory. Synge was the first to investigate the deflection of light rays around a gravitationally intense star \cite{Synge}. Bardeen then argued that the shadow radius of static Schwarzschild black hole is $r_{s}=3M$, and also found that the angular momentum of rotating black holes deformed the shape of the shadow so that it’s not a perfect circle as the static case \cite{Bardeenn}. Moreover, in recent years many studies about the black hole shadow in the context of different space-time geometry have been done \cite{shadow1}--\cite{shadow17}.

On the other hand, it is generally expected that the astrophysical black holes are not in an empty space but instead surrounded by an extremely luminous accretion flow which definitely affects the obtained image of the black hole. There are many papers and interesting discussions, show how the presence of accretion material modifies the black hole shadow. In 1979, Luminet extends the Synge’s work to a more realistic case in which the Schwarzschild black hole is surrounded by an accretion flow \cite{Luminet}. He studied the light deflection of a black hole surrounded by a geometrically thin, optically thick accretion disk and showed that the optical appearance of the black hole and the photon ring depend on both the location and profile of the accretion flow. In an interesting viewpoint, Gralla et al.  by studying the properties of rings around black hole shadow of M87* showed that depending on the intersection points of light rays with the plane of the disk there exist different rings called direct emission, lensing ring and photon ring. In this model, the accretion flow around the black hole is optically thin \cite{Wald}. Moreover, the shadow of the Schwarzschild black hole surrounded by a geometrically thick, optically thin accretion flow has been studied by Cunha et al. in \cite{Cunha2}. A simple spherical model of optically thin accretion in the space-time of the Schwarzschild black hole, and the study of properties of its shadow have been done in \cite{Narayan}. Narayan et al. argue that the behavior of the black hole image in spherical model is different from geometrically thin disk model, where the inner edge of the disk can leave an remarkable imprint on the image, especially when the edge of disk is out of the photon orbit. The shadow of quintessence Schwarzschild black hole surrounded by accretion flow for two types of the accretion flow, optically and geometrically thin disk accretion flow, and also spherically accretion flow, is studied in \cite{Zeng1}. Authors have studied the effects of quintessence state parameter and accretion flow on the optical appearance of black hole \cite{Zeng1}. Also, authors have studied the similar behavior for 4-dimensional Gauss-Bonnet black holes and explored the influence of both the Gauss-Bonnet parameter and spherical accretion flow on the properties of the black hole shadow \cite{Zeng2}. The optical appearance of black holes in the context of Rastall gravity with different spherical accretions has been also studied in Ref. \cite{10}. For further studies about the effect of the location and profile of accretion flow on the black hole shadow in alternative gravity theories see, \cite{0}--\cite{23}.

The astronomical observations show that our universe is in an accelerated expansion phase caused by unknown component with negative pressure and positive energy density, is called dark energy \cite{acc1}--\cite{acc2}. One candidate interpret the negative pressure is the quintessence dark energy. The state equation of quintessence dark energy is $p=\omega\rho$ with $\omega$ being the quintessence state parameter in the range $-1<\omega<-1/3$ \cite{quin1}--\cite{quin2}. The black hole solutions in the presence of quintessence dark energy have been extensively studied. The first static and spherically symmetric black hole solution containing the quintessence matter is obtained by Kiselev \cite{Kiselev1}--\cite{Kiselev2}.

In addition to irregular black holes which have intrinsic singularity in the origin of space-time, black hole solutions without space-time singularity were first introduced by Bardeen \cite{Bardeen}--\cite{Ayon} and later another type of regular black holes was presented by Hayward \cite{Hayward}. Rotating Hayward and charged Hayward black holes are also constructed in \cite{bambi} and \cite{Frolov}. Similar to irregular black holes in the presence of quintessence matter, static Hayward black hole solution with quintessence matter has been obtained in \cite{Pedraza}. The geodesic motion in the space-time of such a black hole is studied in \cite{Pedraza}--\cite{Shil} and extended to the rotating case in \cite{Benavides}. Also, for studying the shadow of rotating Hayward black holes in the absence and presence of quintessence matter see, \cite{He}--\cite{Stuchlik}. However, the influence of quintessence matter on shadows and rings of Hayward black holes surrounded by spherical accretion flow has not yet been studied. So in the present work, similar to Ref. \cite{Zeng1} we study the shadows and rings of Hayward black holes surrounded by spherical accretion, where the effects of quintessence state parameter on the optical appearance of black hole are investigated. We discuss the specific intensity of the shadow and the luminosities of the quintessence Hayward black hole shadow and photon ring, and also compare our results with the results corresponding to the Schwarzschild, quintessence Schwarzschild, and Hayward black holes.

The paper is structured as follows. In section \ref{1-Hayward black holes with quintessence matter}, we present a brief review of Hayward black holes surrounded by quintessence matter and some of their properties. In section\ref{1-Light deflection by the quintessence Hayward black hole} we discuss the photon trajectories in the space-time of quintessence Hayward black holes and investigate the effect of the model parameters on them. In section \ref{1-Shadows and photon rings with spherical accretion flow} we present the shadow images of quintessence Hayward black hole with static and spherical accretion flow. The paper ends with drawing conclusions.

\section{Hayward black holes with quintessence matter}
\label{1-Hayward black holes with quintessence matter}
The geometry of a Hayward black hole surrounded by quintessence dark energy can be expressed as \cite{Pedraza}
\begin{equation}
ds^2=-f(r)dt^2+\frac{dr^2}{f(r)}+r^2\left(d\theta^2+\sin ^2\theta d\varphi^2\right),
\label{1}
\end{equation}
where
\begin{equation}
f(r) = 1-\frac{2 M r^2}{r^3+2 M\epsilon^2}-\frac{a}{r^{3 \omega+1}},
\label{2}
\end{equation}
that $M$ is the mass of black hole, $\epsilon$ is a parameter related to the cosmological constant and $a$ and $\omega$ are the normalization factor and the state parameter of the quintessence matter, respectively. In Ref.\cite{Pedraza} authors express the black hole mass, the
radial distance and the parameter $a$ in units of the $\epsilon$, but here we define $2 M\epsilon^2 \equiv g^3$ and investigate the effect of $g$ parameter in our calculations. In the case of $a = 0$, the above metric reduces to the Hayward black hole obtained by Hayward \cite{Hayward}, and for $\epsilon=0$ represents the Schwarzschild black hole surrounded by quintessence matter that was initially obtained by Kiselve in \cite{Kiselev1}. Also, for $a=\epsilon=0$ behaves like the Schwarzschild black hole that has only one event horizon. We also note by choosing different values of $\omega$, the different black hole solutions can be obtained. For instance, when $\omega=-\frac{1}{3}$ the solution (\ref{1}) corresponds to the Hayward black hole \cite{Hayward}. Moreover, in the case of $\omega=\frac{1}{3}$ and $a=-Q^2$ corresponds to the charged Hayward black hole \cite{Frolov}.

\begin{figure}[H]
\centering
\includegraphics[width=3.0in]{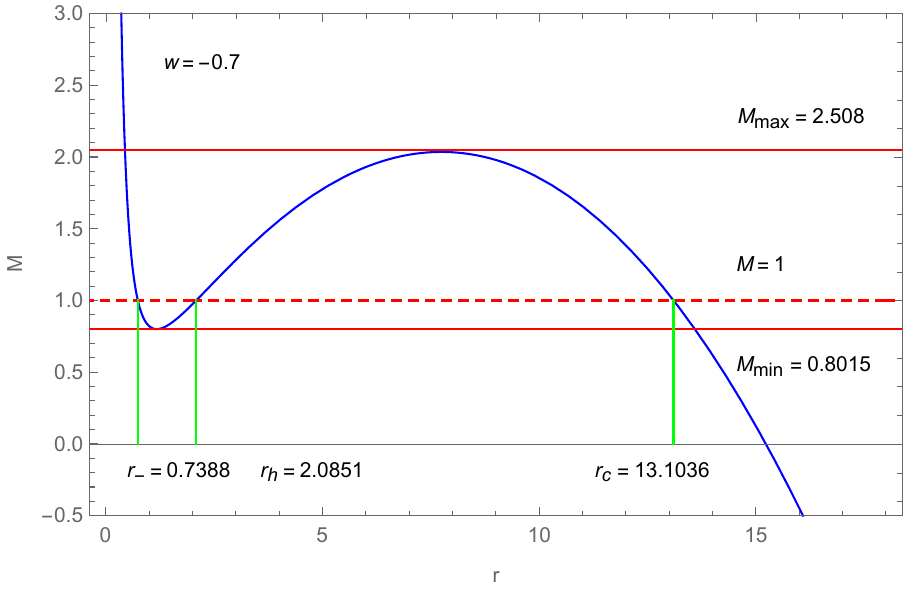}
\caption{The black hole mass as a function of the radial coordinate $r$ for $g = 0.9$, $a = 0.05$ and $\omega = -0.7$.}
\label{Mass-a}
\end{figure}

The behavior of $f(r)$ function and thus the black hole horizons depend on the parameters $g$, $a$ and $\omega$. For quintessence dark energy the state parameter takes the values in the interval $-1<\omega<-\frac{1}{3}$ and the energy density of quintessence matter is given by
\begin{equation}
\rho = -\frac{3 \omega a}{2 r^{3 \omega+1}}.
\label{3}
\end{equation}
As is clear, for a positive energy density the parameter $a$ should be positive. In the present work, we take $\omega = -0.5$ and $\omega = -0.7$ and thus we investigate the behavior of horizons for two cases. The interested reader is referred to \cite{Pedraza} for more discussions about properties of the black hole horizons. In what follows, we find the conditions that the Hayward metric surrounded by quintessence matter could have three horizons. From equation $f(r)=0$ one can obtain the relationship between the black hole mass and the horizon radius as follows
\begin{equation}
M = \frac{r^3+g^3}{2 r^2}(1-\frac{a}{r^{3\omega+1}}).
\label{4}
\end{equation}
Choosing suitable values of the constants $g$, $a$ and $\omega$ the function $M$ is plotted in Fig.~\ref{Mass-a}. It can be seen that the function $M$ has a local maximum and minimum both located above the horizontal axis. For values of $M$ in the interval $M_{\rm min}<M<M_{\rm max}$ the black hole has three horizons. The radii of inner and event horizons are represented by $r_-$ and $r_h$ respectively, while $r_c$ is the cosmological horizon. There exist a critical value of normalization factor, $a$, for which $r_-=r_h=r_c$. This critical value $a_{\rm critical}$ is given by
\begin{equation}
a_{\rm critical} = -\frac{4+3\omega+\sqrt{16+9\omega^2}}{2-3\omega+\sqrt{16+9\omega^2}}\left(\frac{\omega-4-\sqrt{16+9\omega^2}}{2 \omega}\right)^{\omega+1/3}\frac{g^{3\omega+1}}{3\omega},
\label{5}
\end{equation}
when $a>a_{\rm critical}$, there exist no black hole solution for any values of $M$, but in the case of $a<a_{\rm critical}$ the black hole solutions exist for $M_{\rm min}<M<M_{\rm max}$.

In Table~\ref{T1}, we present the values of $M_{\rm min}$, $M_{\rm max}$ for $a=0.05$, $\omega=-0.7$ and different values of $g$. Also from equation (\ref{5}) the value of $a_{\rm critical}$ is obtained for $\omega=-0.7$ and different values of $g$. In this work, we take $a=0.05$, and $M=1$ which is in interval $M_{\rm min}$ and $M_{\rm max}$. In this case the value of $a$ is smaller than $a_{\rm critical}$ for all values of $g$ and thus we have the black hole solutions with three horizons.

\begin{table}[H]
\centering
\caption{\footnotesize The value of $a_{\rm critical}$ and also the values of $M_{\rm min}$, $M_{\rm max}$ for $a = 0.05$, $\omega=-0.7$ and different values of $g$.}
\begin{tabular}{l l l l l l l l}
\hline
$\omega$&$g$&$a_{\rm critical}$&$M_{\rm min}$&$M_{\rm max}$\\ [0.5ex]
\hline\\
    {- 0.7} &0.1&1.81558&0.47548&2.03223\\
    {- 0.7} &0.3&0.54223&0.48783&2.03234\\
   {- 0.7}  &0.5&0.30914&0.45812&2.03277\\
    {- 0.7} &0.7&0.21351&0.63224&2.03372 \\
   {- 0.7} &0.9&0.16194& 0.80067 &2.03540\\
\hline\\
 \label{T1}
\end{tabular}
\end{table}

\begin{figure}[H]
\centering
\includegraphics[width=3.0in]{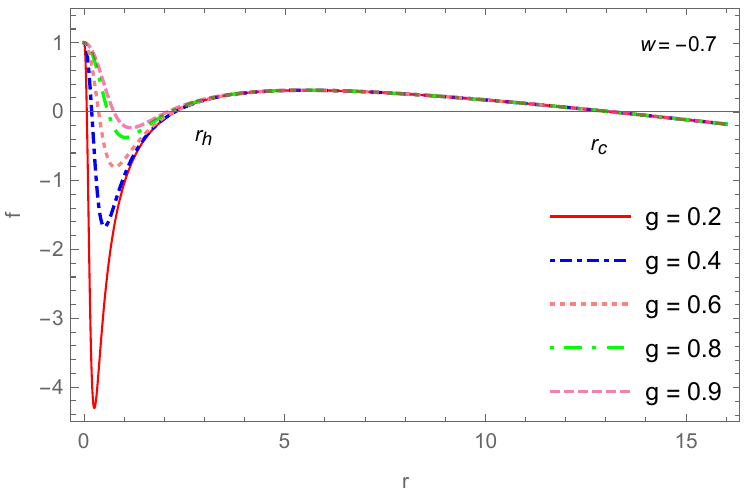}
\caption{The metric coefficient $f(r)$ as a function of the radial coordinate $r$ for $M = 1$, $a = 0.05$ and $\omega = -0.7$ with different values of $g$.}
\label{event-horizon}
\end{figure}

The behavior of the function $f(r)$ for $M = 1$, $a = 0.05$ and $\omega = -0.7$ is plotted in Fig.~\ref{event-horizon}. As the figure shows, $f(r)$ function has three real roots corresponding to $r_-$, $r_h$ and $r_c$, respectively \cite{Pedraza}. Also the difference between event horizon and the cosmic horizon decreases with the increases of absolute value of $\omega$ parameter, see Table~\ref{T2}.

\section{Light deflection by the quintessence Hayward black hole}
\label{1-Light deflection by the quintessence Hayward black hole}
To obtain the light deflection of quintessence Hayward black holes we study the null geodesics in the space-time of such black holes. Thus, we consider the Euler-Lagrange equation as follows
\begin{equation}
\frac{d}{ds}\left(\frac{\partial {\cal L}}{\partial {\dot{x}^{\mu}}}\right)-\frac{d{\cal L}}{dx^{\mu}}= 0,
\label{6}
\end{equation}
where the Lagrangian density ${\cal L}$ is given by
\begin{equation}
{\cal L}  = \frac{1}{2}g_{\mu\nu}\dot{x}^{\mu}\dot{x}^{\nu} = \frac{1}{2}\left(-f(r) \dot{t}^2+\frac{1}{f(r)} \dot{r}^2 + r^2 \dot{\theta}^2+ r^2 \sin^2{\theta}\dot{\varphi}^2\right),
\label{7}
\end{equation}
$\dot{x}^{\mu}$ is the photon four-velocity, i.e., $\dot{x}^{\mu}=\frac{\partial x^{\mu}}{\partial s}$ where $s$ denotes the affine parameter. For a spherically symmetric space-time without loss of generality, we consider the motion in the equatorial plane and impose two conditions $\theta=\frac{\pi}{2}$ and $\dot{\theta}=0$ . Also, since the metric coefficients have no explicit dependence on $t$ and $\varphi$ coordinates, there are two conserved quantities corresponding to the energy $E$ and angular momentum $L$, of the photon. Now, using equations (\ref{2}), (\ref{6}) and (\ref{7}) one can find
\begin{equation}
\dot{t} = -\frac{E}{1-\frac{2 M r^2}{r^3+g^3}-\frac{a}{r^{3\omega+1}}},
\label{8}
\end{equation}
\begin{equation}
\dot{\varphi} = \frac{L}{r^2},
\label{9}
\end{equation}
\begin{equation}
\dot{r}^2+\left(1-\frac{2 M r^2}{r^3+g^3}-\frac{a}{r^{3 \omega+1}}\right)\left(\frac{L^2}{r^2}+h\right) = E^2,
\label{10}
\end{equation}
where $h=0$ and $h=1$ correspond to the null-like and time-like geodesics, respectively. Considering the null geodesics with $h=0$ and using equation (\ref{10}), the effective potential can be rewrite as
\begin{equation}
V_{\rm eff} (r) = \frac{ L^2}{r^2}f(r) = \frac{ L^2}{r^2}\left(1-\frac{2 M r^2}{r^3+g^3}-\frac{a}{r^{3\omega+1}}\right).
\label{12}
\end{equation}

\begin{figure}[H]
\centering
\includegraphics[width=3.0in]{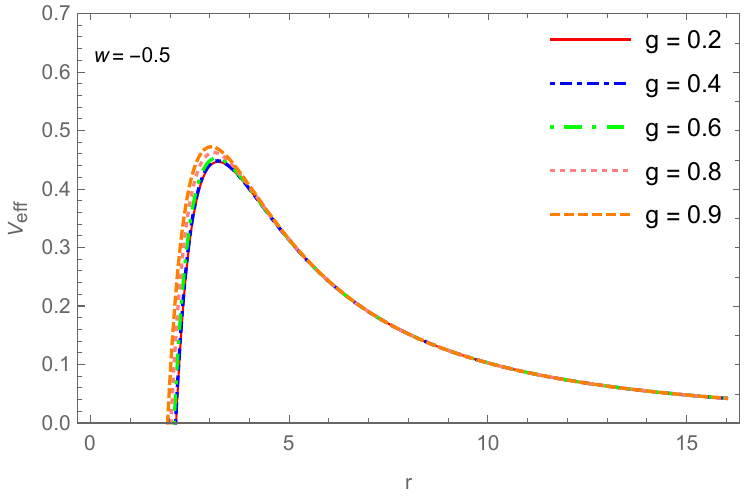}
\includegraphics[width=3.0in]{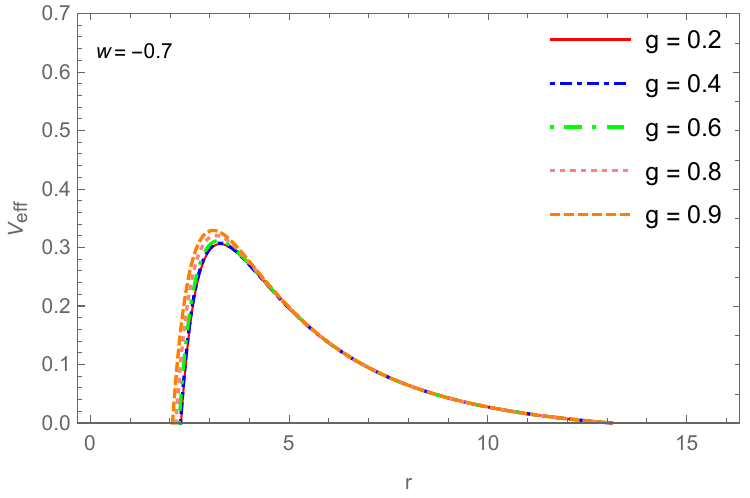}
\caption{ The effective potential as a function of the radial coordinate $r$  for different values of $g$ with $\omega = -0.5$ (left panel) and $\omega = -0.7$ (right panel) and $a = 0.05$.}
\label{effetive-potential-1}
\end{figure}

\begin{figure}[H]
\centering
\includegraphics[width=3.0in]{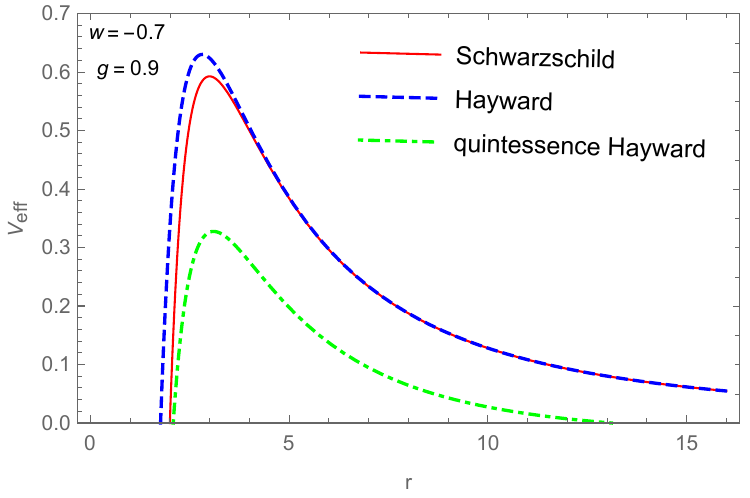}
\caption{The effective potential as a function of the radial coordinate $r$  for Schwarzschild space-time $g = a =0$, Hayward black hole for $g = 0.9$, $a = 0$, and Hayward black hole surrounded by quintessence matter with $g = 0.9$, $a = 0.05$, $w = -0.7$ and $M = 1$.}
\label{effetive-potential-2}
\end{figure}

In Fig.~\ref{effetive-potential-1}, the effective potential of quintessence Hayward black holes is plotted for different values of the parameter $g$. The state parameter set to $\omega=-0.5$ and $\omega=-0.7$ in the left and right panels, respectively. It can be seen that increasing $g$ leads to an increase in the peak of the potential, while with increasing the absolute value of $\omega$ the maximum of the potential decreases. The effect of quintessence parameter on the effective potential is displayed in Fig.~\ref{effetive-potential-2}, showing that in the presence of quintessence dark energy the maximum of the effective potential decreases, and for all values of $g$ the peak of the potential for quintessence Hayward black holes is less than that for Hayward and Schwarzschild black holes.

\begin{figure}[H]
\centering
\includegraphics[width=3.0in]{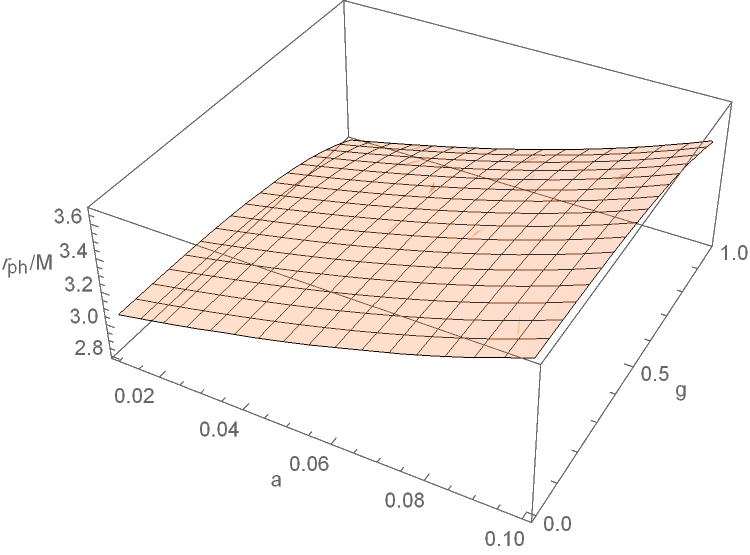}
\caption{The behavior of the photon radius as a function of the parameters $g$ and $a$ for $\omega= -\frac{2}{3}$.}
\label{photon-radius}
\end{figure}

Now, we are going to obtain the photon sphere and impact parameter of photon sphere for quintessence Hayward black holes by employing the ray-tracing method. The photon orbit occurs at $r=r_{ph}$ and thus from equation (\ref{10}) the conditions for such orbits are given by
\begin{equation}
V_{\rm eff}(r_{ph}) = E_{ph}^2,\hspace {0.5 cm} V^{'}_{\rm eff}(r_{ph}) = 0,
\label{13}
\end{equation}
where the prime represents differentiation with respect to the radial coordinate $r$. Use of equation (\ref{12}) leads to the following relation
\begin{equation}
r {f^{'}}(r)-2f(r)=0.
\label{14}
\end{equation}
Substituting $f(r)$ from equation (\ref{2}) we obtain the following equation for the photon sphere
\begin{equation}
r^6 -3 m r^5 -4 g^3 r^3 -2g^6 +3 a (1+\omega) (g^3+r^3)^2 r^{-(3\omega+1)} = 0,
\label{15}
\end{equation}
which does not have an analytical solution and thus we numerically obtain the roots of that. For $a=g=0$, we find $r_{ph}=3M$ which is the photon radius for the Schwarzschild space-time. Also, in the case of $a=0, g\neq0$ namely for the Hayward black holes the above equation does not have an analytical solution while, for the Schwarzschild black holes in the presence of quintessence matter with $a\neq0$ and $g=0$ there is only an exact solution for $\omega=-\frac{2}{3}$ as follows
\begin{equation}
r_{ph} = \frac{1-\sqrt{1-6 M a}}{a}.
\label{16}
\end{equation}
Moreover, the impact parameter of the photon sphere defined as $b_{ph}=L_{ph}/E_{ph}$, is given by \cite{Bambi}
\begin{equation}
b_{ph} = \frac{r_{ph}}{\sqrt{f(r_{ph})}}.
\label{17}
\end{equation}
The numerical results of the inner horizon radius $r_-$, event horizon $r_h$, cosmological horizon $r_c$, the radius of the photon sphere $r_{ph}$ and the impact parameter of photon sphere $b_{ph}$ for different values of $g$ and $\omega$ with $M=1$ and $a=0.05$ are presented in Table~\ref{T2}. Cosmological horizon has an important role in the observed shadow of quintessence Hayward black holes, since we consider the distant observer nearby the cosmological horizon and the accretion flow near the event horizon.  The difference between event horizon and cosmological horizon depends on quintessence state parameter $\omega$ and with increasing absolute value of $\omega$, this difference decreases. When we fixed the value of $\omega$ and increased the $g$ parameter, $r_h$, $r_c$, $r_{ph}$ and $b_{ph}$ decrease, however the rate of this decrease is very small for cosmic horizon $r_c$. We also see that for a fixed $g$ with increasing the absolute value of $\omega$ the values of $r_h$ and $b_{ph}$ increase but the cosmological horizon $r_c$ decreases. Moreover, the photon radius $r_{ph}$ increases and then decreases around $\omega=-0.7$.

In Table~\ref{T3} , we compare the numerical values of $r_h$, $r_c$, $r_{ph}$ and $b_{ph}$ for the Schwarzschild space-time, Hayward black hole, quintessence Schwarzschild black hole and quintessence Hayward black hole. The first row corresponds to the results of Schwarzschild black hole and the second row to those of Hayward regular black hole. The third and fourth rows correspond to the Schwarzschild and Hayward black holes surrounded by quintessence, respectively. The results of Table~\ref{T3} on scales of the size of the horizons, photon sphere and impact parameter show that Hayward and quintessence Hayward black holes with a regular core do not significantly different from irregular Schwarzschild and quintessence Schwarzschild black hole, and seems unnecessary when other, more important ingredients such as quintessence background and accretion flow are present and presumably dominate.

For quintessence Hayward black holes under consideration, the dependence of the photon radius on the parameters $a$ and $g$ with $\omega=-\frac{2}{3}$ is plotted in Fig.~\ref{photon-radius}. As the figure shows, for a fixed value of $a$ with increasing $g$ the photon radius decreases, while at fixed $g$ by increasing $a$ the photon radius increases. Also, it can be seen that for $a=g=0$, $r_{ph}=3M$ which is the radius of the photon sphere for the Schwarzschild black hole.

\begin{table}[H]
\centering
\caption{\footnotesize  The values of inner horizon, $r_-$, event horizon, $r_h$, cosmological horizon, $r_c$, photon radius, $r_{ph}$, and impact parameter, $b_{ph}$, for different values of $g$ and $\omega$, with $M =1$, $a=0.05$. The results in the first row of each $\omega$ corresponds to the quintessence schwarzschild black hole.}
\begin{tabular}{l l l l l l l l}
\hline
$\omega$&$g$&$r_{-}/M$& $r_{h}/M$&$r_c/M$&$r_{ph}/M$&$b_{ph}/M$\\ [0.5ex]
\hline
{- 0.4}  &0     &$-$   &2.1234   &319999  &3.1804    &5.7295\\
     &0.1     &0.0222   &2.1232  &319999      &3.1802    &5.7293\\
    &0.3     &0.1177   &2.1173   &319999     &3.1749   &5.7246\\
      &0.5     &0.2623   &2.0946   &319999      &3.1549    &5.7064\\
      &0.7     &0.4586   &2.0399  &319999      &3.1081   &5.6646\\
      &0.9     &0.7298   &1.9239   &319999     &3.0159    &5.5844\\
  \\
{- 0.5}  &0     &$-$   &2.1586   &395.969            &3.2163    &5.9881\\
      &0.1     &0.0224   &2.1584   &395.969      &3.2161    &5.9879\\
     &0.3     &0.1187   &2.1525   &395.969     &3.2109   &5.9828\\
      &0.5     &0.2644   &2.1299   &395.969      &3.1908    &5.9635\\
      &0.7     &0.4618   &2.0756   &395.969      &3.1439    &5.9189\\
      &0.9     &0.7331   &1.9610   &395.969     &3.0518    &5.8335\\
\\
{- 0.6}  &0     &$-$   &2.2080   &39.6448          &3.2504    &6.4204\\
      &0.1     &0.0224   &2.2078   &39.6448       &3.2502    &6.4201\\
     &0.3     &0.1192   &2.2019   &39.6448       &3.2449    &6.4142\\
      &0.5     &0.2658   &2.1794   &39.6448       &3.2246    &6.3919\\
      &0.7     &0.4643   &2.1252   &39.6448      &3.1772   &6.3403\\
      &0.9     &0.7361   &2.0115   &39.6448      &3.0841    &6.2498\\
\\
{- 0.7} &0     &$-$      &2.2830   &13.1027  &3.2710    &7.2343\\
      &0.1     &0.2246   &2.2828   &13.1027      &3.2708    &7.2340\\
      &0.3     &0.1194   &2.2769   &13.1028      &3.2654    &7.2259\\
     &0.5     &0.2665   &2.2539  &13.1029     &3.2446    &7.1952\\
     &0.7     &0.4656   &2.1992   &13.1031     &3.1961    &7.1245\\
      &0.9     &0.7379   &2.0851   & 13.1036     &3.1007    &6.9900\\
\\

{- 0.8} &0     &$-$   &2.4149        &6.5509  &3.2548   &9.2206\\
    &0.1     &0.0225   &2.4146   &6.5509     &3.2546    &9.2199\\
      &0.3     &0.1198   &2.4083   &6.5513    &3.2489    &9.2027\\
     &0.5     &0.2674   &2.3839   &6.5523     &3.2275    &9.1375\\
      &0.7     &0.4678   &2.3262   &6.5547     &3.1773    &8.9892\\
      &0.9     &0.7413   &2.2075   &6.5587     &3.0784    &8.7135\\
\hline
\end{tabular}
\label{T2}
\end{table}

\begin{table}[H]
\centering
\caption{\footnotesize  The values of event horizon, $r_h$, cosmological horizon, $r_c$, photon radius, $r_{ph}$, and impact parameter, $b_{ph}$, for Schwarzschild space-time, Hayward black hole, Schwarzschild and Hayward black holes surrounded by quintessence matter.}
\begin{tabular}{l l l l l l l l}
\hline
$Type$&$\omega $&$a$&$g$& $r_{h}/M$&$r_c/M$&$r_{ph}/M$&$b_{ph}/M$\\ [0.5ex]
\hline
 {Schwarzschild}      &   $-$ &0 &0              &2         &$- $    &3    &5.19615\\
{Hayward}      &$-$  &0 &0.5            &1.96772   &$-$      &2.97162    &5.17169\\
\\
{Quintessence Schwarzschild}      &   - 0.5  &0.05  &0             &2.15857   &395.9690      &3.21631    &5.98805\\
{Quintessence Hayward}        & - 0.5 &0.05   &0.5            &2.12987   &395.9690      &3.19078    &5.96347\\
\hline
\end{tabular}
\label{T3}
\end{table}

\begin{figure}[H]
\centering
\includegraphics[width=3.0in]{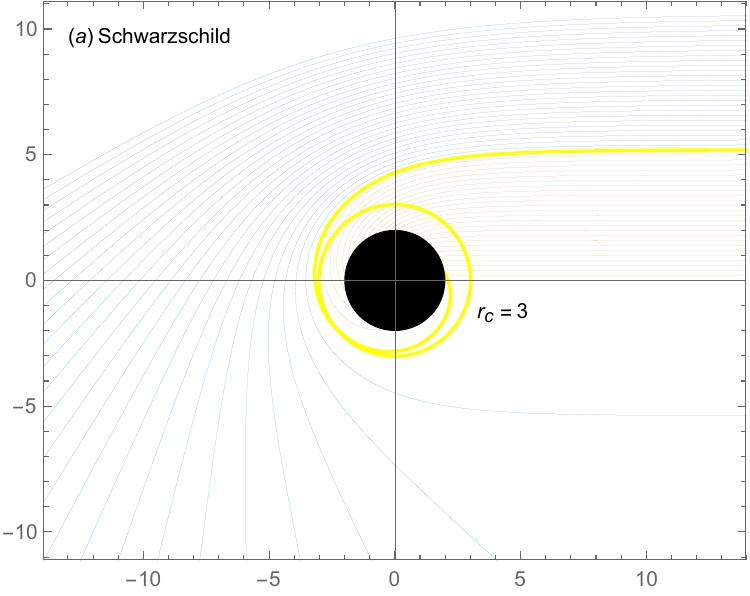}
\includegraphics[width=3.0in]{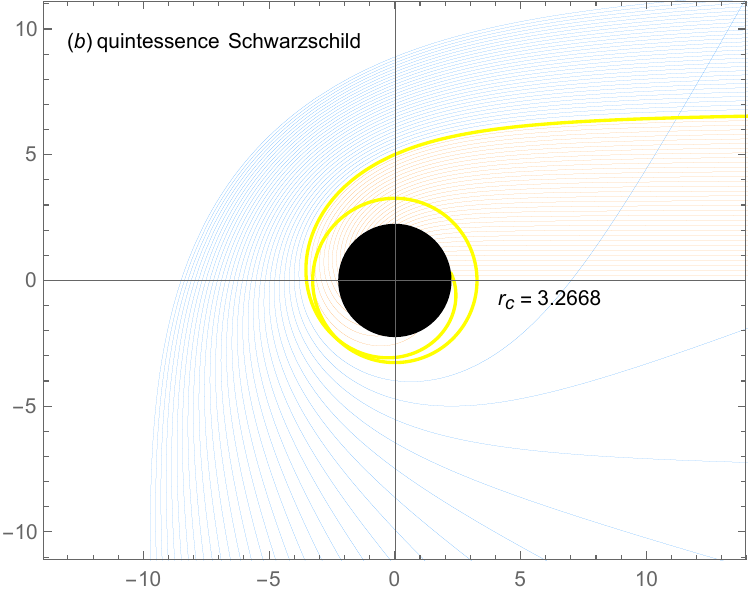}
\includegraphics[width=3.0in]{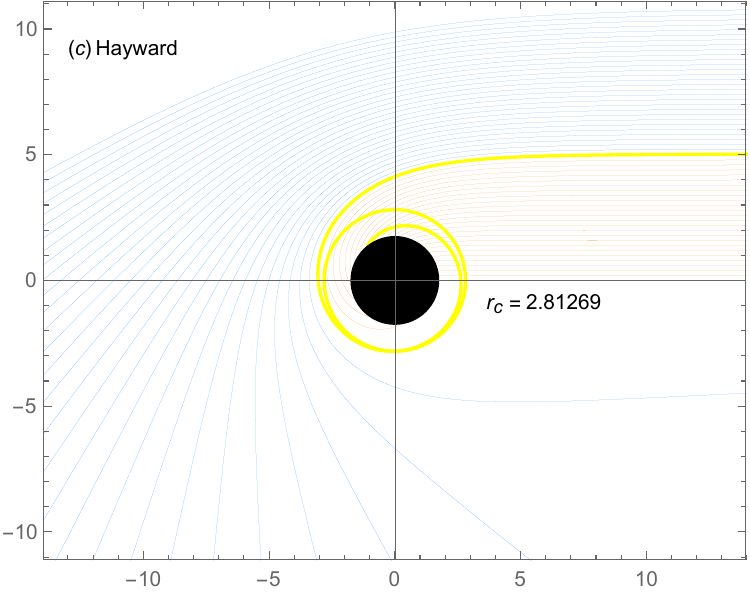}
\includegraphics[width=3.0in]{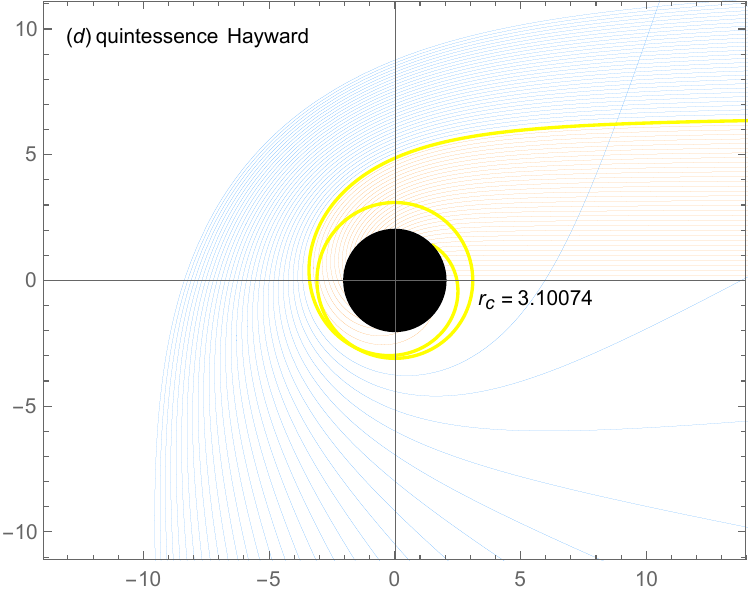}
\caption{The polar plot of light rays around a schwarzschild black hole with and without quintessence matter (top row) and a Hayward black hole with and without quintessence matter (bottom row) for $g=0.9$ and $a=0.05$, $\omega=-0.7$ with $M =1$ . The yellow, blue and orange curves correspond to the trajectory of light rays with $E = E_{ph}$, $E < E_{ph}$ and $E > E_{ph}$, respectively.}
\label{light-trajectory}
\end{figure}

Using equations (\ref{9}) and (\ref{10}) and new variable $u=\frac{1}{r}$ we find
\begin{equation}
\frac{du}{d\varphi} = \sqrt{\frac{1}{b^2}-u^2 \left(1-\frac{2 M u}{1+g^3 u^3}-au^{3\omega+1}\right)}.
\label{18}
\end{equation}

Assuming that light rays approach the black hole from the right side, shadow and photon sphere are formed from the deflection of light corresponds to the Fig.~\ref{light-trajectory}. The light trajectories in the space-time of Schwarzschild, quintessence Schwarzschild, Hayward and quintessence Hayward black holes are plotted in panels (a) to (d). As can be seen, by increasing the parameter $g$ the shadow radius decreases, while in the presence of quintessence matter an enhancement in the absolute value of $\omega$, increases the shadow radius which is in agreement with Table~\ref{T2}. The top row shows the results of the Schwarzschild and quintessence Schwarzschild black holes while the bottom row shows the results of regular Hayward black holes in the absence and presence of quintessence matter. By comparing figures (a) with (c) and also (b) with (d), we find that by increasing the parameter $g$ the light deflection is smaller, namely the presence of the parameter $g$ cause to decrease the strength of the gravitational field and thus diminishes the light bending angle \cite{Jusufi}. On the other hand, comparing the results of figures (a) with (b) and also results of (c) with (d) show that the presence of quintessence matter increases the deflection of light rays \cite{Javed}.

Since the shadow is the fingerprint of the metric of space-time, one can constrain the parameter of the space-time geometry by the observed shadow \cite{n0}-\cite{n7}. The angular diameter $\Omega$ of the black hole shadow for a distant observer can be defined as \cite{n0}
\begin{equation}
\Omega=\frac{2 b_{ph}}{D},
\label{n1}
\end{equation}
where $D$ is the distance between the black hole and distant observer. Equation (\ref{n1}) can be rewritten as
\begin{equation}
\left(\frac{\Omega}{\mu as}\right) = \left(\frac{6.191165\times10^{-8}}{\pi}\frac{\gamma}{D/Mpc}\right)\left(\frac{b_{ph}}{M}\right),
\label{n2}
\end{equation}
here  $\gamma$ is the mass ratio of the black hole to the Sun and the impact parameter of the photon sphere, $b_{ph}$, is obtained from equation (\ref{17}).

\begin{figure}[H]
\centering
\includegraphics[width=3.0in]{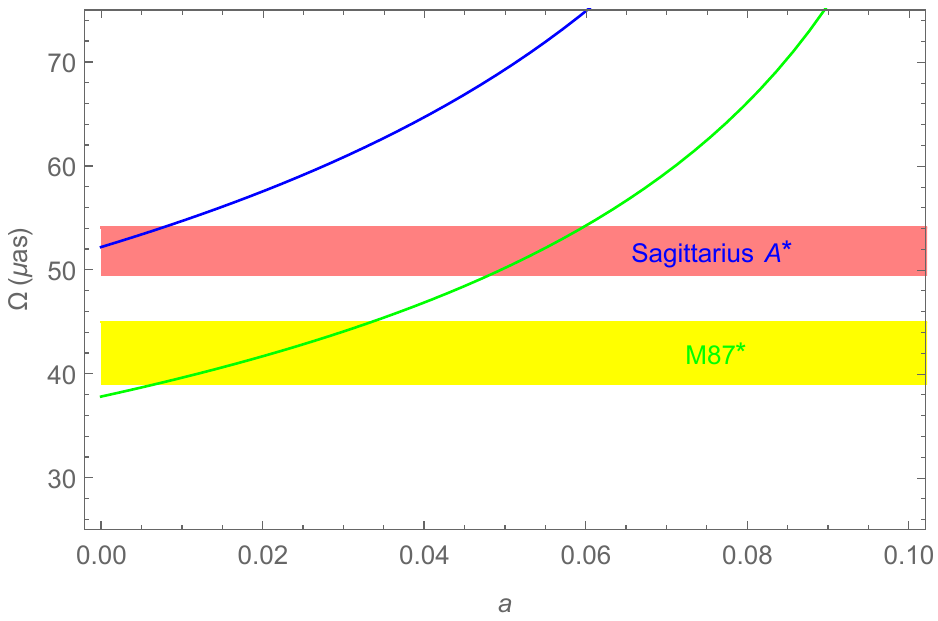}
\caption{ The relation between the angular diameter $\Omega$ of the observed shadow and the parameter $a$ for a quintessence Schwarzschild black hole with $w=-2/3$. The pink and yellow regions are the experimental data of Sagittarius A* ($51.8 \pm 2.3$ $\mu as$) and M87* ($42 \pm 3$ $\mu as$) and reported by the EHT, respectively.}
\label{new}
\end{figure}

For quintessence Hayward black hole, one can obtain the constraints of the free parameters using the shadow diameter estimated by the EHT observations. We
consider the quintessence Schwarzschild black hole ($g=0$) with $\omega = -2/3$  and constrain parameter $a$ for both M87* and Sagittarius A* in  Fig. \ref{new}. Current observations correspond to $\gamma = 4.14\times10^{6}$, and distance $D = 8.127$ kpc for Sagittarius A* and $\gamma = 6.2\times10^{9}$, and distance $D = 16.8$ Mpc for M87*, respectively \cite{A1}-\cite{A7}. The yellow and pink regions are the shadow diameters of M87* ($42 \pm 3$ $\mu as$) and Sagittarius A* ($51.8 \pm 2.3$ $\mu as$) reported by the EHT observations. Consequently, we constrain the parameter $a$ of the quintessence Schwarzschild black hole as $ 0.007 < a < 0.033$ for M87*, and $0 < a < 0.0078$ for Sagittarius A*.

\section{Shadows and photon rings with spherical accretion flow}
\label{1-Shadows and photon rings with spherical accretion flow}
Now, we are going to study the optical appearance of Hayward black hole surrounded by the quintessence matter with the spherical accretion flow and investigate the influence of the parameter $g$, and quintessence state parameter, $\omega$, on it. First, we consider a static spherical accretion model of optically and geometrically thin accretion flows and then focus on infalling spherical accretion flows.
\subsection{The static spherical accretion}
First we investigate the shadow and photon ring in the background of a quintessence Hayward black hole surrounded by static spherical accretion flow. The observed specific intensity can be obtained by integrating the specific emissivity along any ray path \cite{flow1}--\cite{flow2}
\begin{equation}
I(\nu_{\rm obs}) = \int_{\gamma} g^3 j(\nu_{\rm em}) dl_p,
\label{19}
\end{equation}
with
\begin{equation}
g \equiv \frac{\nu_{\rm obs}}{\nu_{\rm em}}.
\label{20}
\end{equation}
In the above equation, $g$ is the redshift factor, while $\nu_{\rm obs}$ and $\nu_{\rm em}$ denote the observed photon frequency and radiated photon frequency, respectively. In the quintessence Hayward black hole space-time (\ref{1}), the redshift factor is $g = f(r)^{1/2}$. Assuming that the emission is monochromatic with rest frame frequency $\nu_{\rm t}$ and the emission radial profile is $1/r^2$ \cite{flow2}, the specific emissivity takes the form
\begin{equation}
j (\nu_{\rm em}) \propto \frac{\delta(\nu_{\rm em}-\nu_{t})}{r^2}.
\label{22}
\end{equation}
Also, according to equation (\ref{1}), the proper length measured in the rest frame of the emitter is given by
\begin{equation}
dl_p = \sqrt{\frac{1}{f(r)}dr^2+r^2 d\varphi^2} = \sqrt{\frac{1}{f(r)}+r^2 \left(\frac{d\varphi}{dr}\right)^2} dr.
\label{23}
\end{equation}
Thus, by substituting equation (\ref{18}) into equation (\ref{23}), we obtain the specific intensity observed by a distant observer as
\begin{equation}
I (\nu_{\rm obs}) = \int_{\gamma}\frac{f(r)}{r^2}\sqrt{1+\frac{b^2 f(r)}{r^2-b^2 f(r)} } dr.
\label{24}
\end{equation}

\begin{figure}[H]
\centering
\includegraphics[width=3.0in]{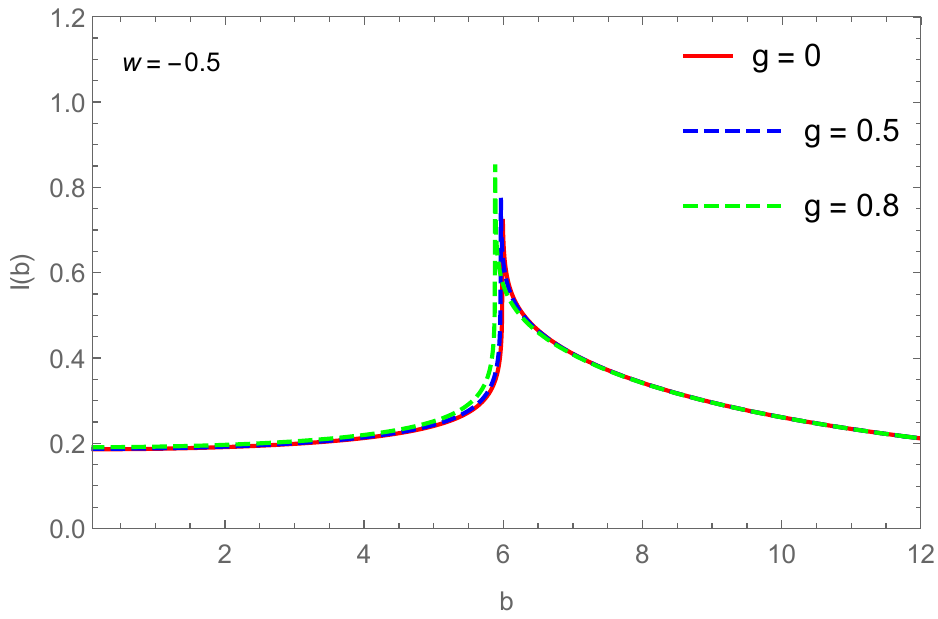}
\includegraphics[width=3.0in]{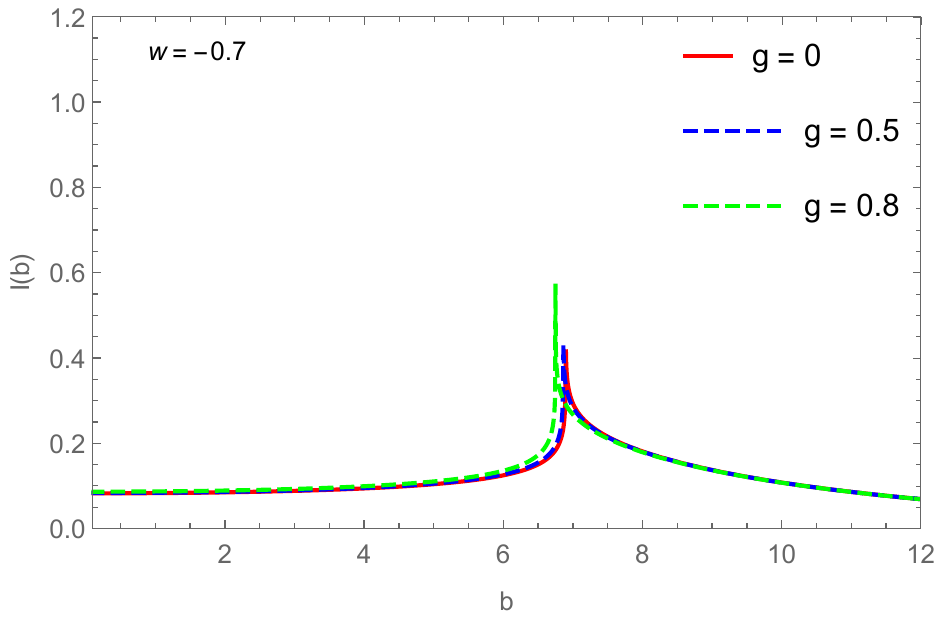}
\caption{ The observed intensity $I(\nu_{\rm obs})$ for static spherical accretion flow around a quintessence Hayward black hole $\omega = -0.5$ (left panel) and $\omega = -0.7$ (right panel) for different value of $g$, $a = 0.05$ and $M = 1$.}
\label{intensity-static}
\end{figure}

From equation (\ref{18}), it is clear that the trajectory of light rays depends on impact parameter $b$, so that in the case of $b=b_{ph}$, ($E=E_{ph}$), the photons will revolve around the black hole several times and for $b>b_{ph}$, ($E<E_{ph}$), the light rays deflect by it. Moreover, the photons with $b<b_{ph}$, ($E>E_{ph}$), eventually fall into the black hole singularity. Therefore, in order to study the photon intensity which associate with the light trajectories, we plotted the observed intensity as a function of impact parameter $b$. In Fig.~\ref{intensity-static}, the plots of intensity as a function of $b$ for different values of the parameter $g$, with $\omega=-0.5$ (left panel) and $\omega=-0.7$ (right panel) are presented. Clearly, the intensity increases with impact parameter, sharply peak at $b=b_{ph}$ and then decreases with $b$. The reason for maximum intensity at $b=b_{ph}$ is that the photon revolves around the black hole several times as it approaches the photon sphere. Meanwhile, at fixed $\omega$ the peak value of intensity increases with increasing $g$ so that the quintessence Schwarzschild black hole with $g=0$ has the smallest value. Also, for a given value of $g$ the larger absolute value of $\omega$ has the smaller intensity, namely the case of $\omega=-0.7$ has the smaller intensity in comparison with $\omega=-0.5$. It shows that in the presence of quintessence dark energy the observed intensity decreases.

\begin{figure}[H]
\centering
\includegraphics[width=1.8in]{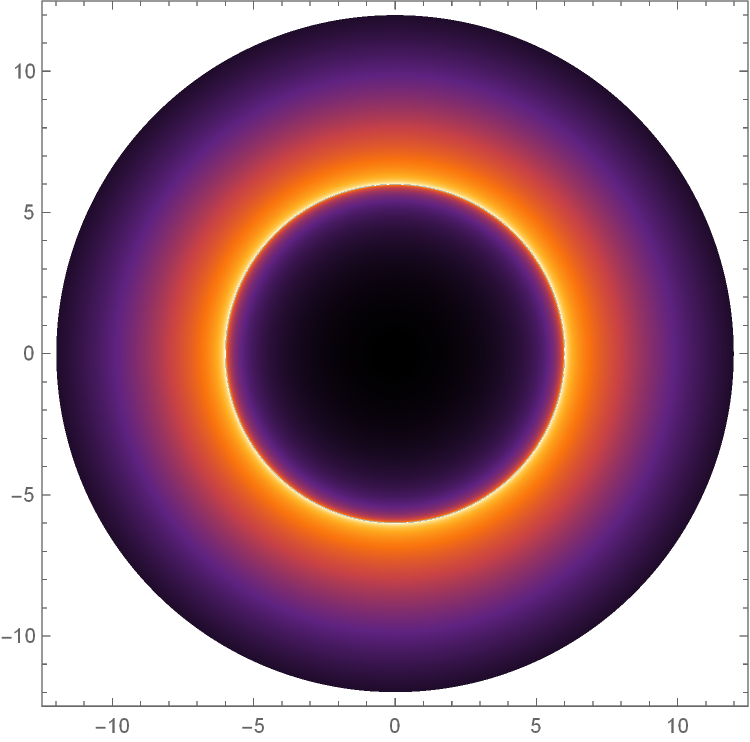}
\includegraphics[width=0.2in]{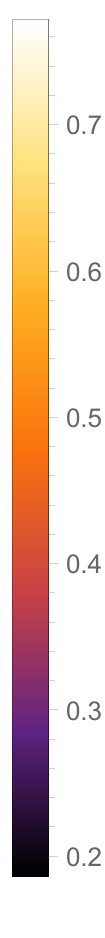}
\includegraphics[width=1.8in]{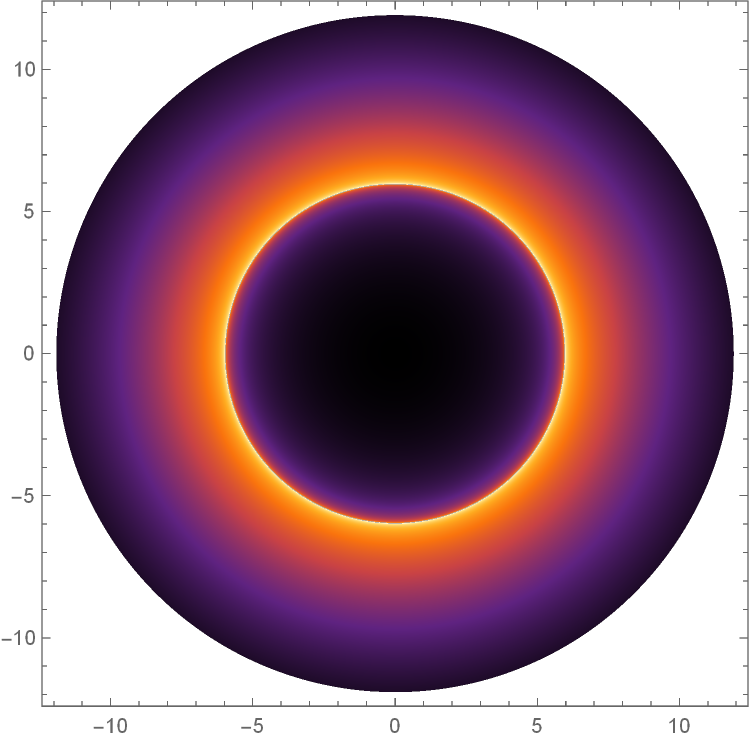}
\includegraphics[width=0.2in]{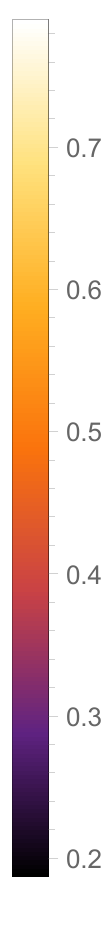}
\includegraphics[width=1.8in]{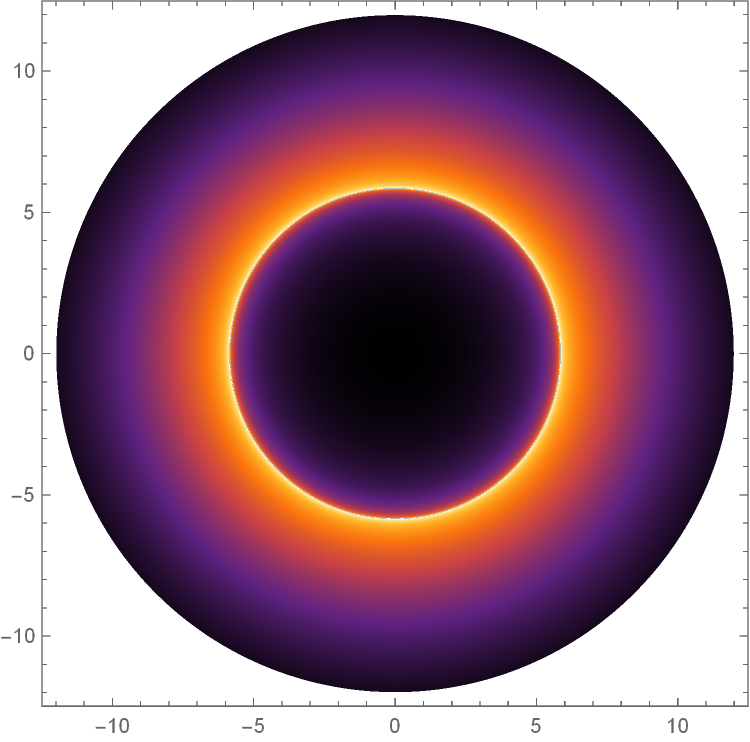}
\includegraphics[width=0.2in]{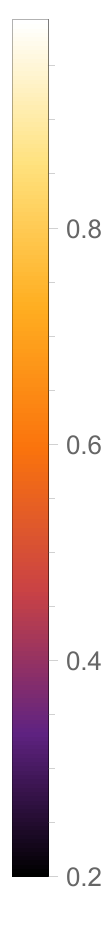}
\includegraphics[width=1.8in]{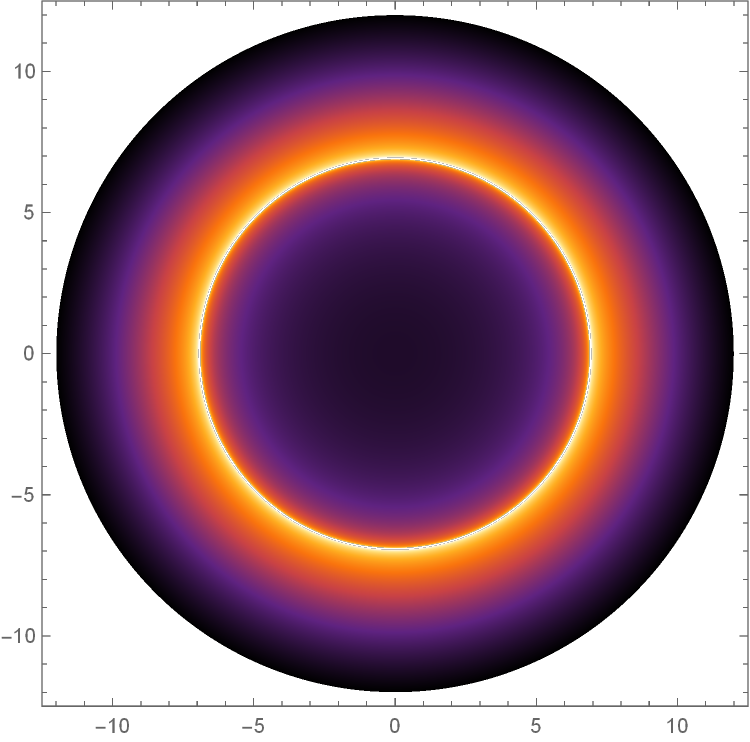}
\includegraphics[width=0.2in]{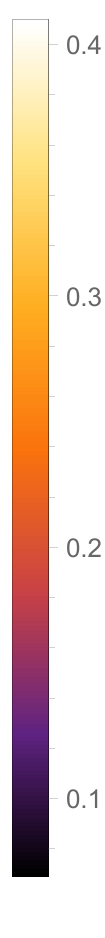}
\includegraphics[width=1.8in]{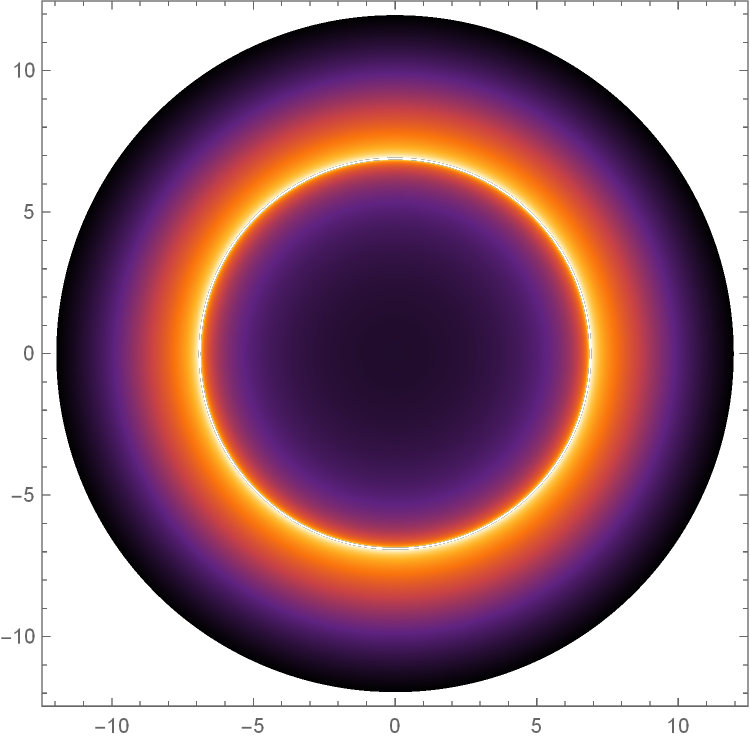}
\includegraphics[width=0.2in]{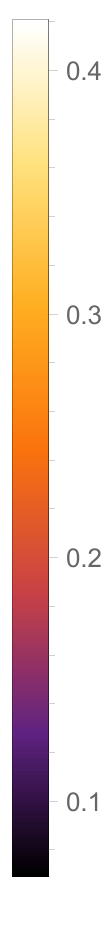}
\includegraphics[width=1.8in]{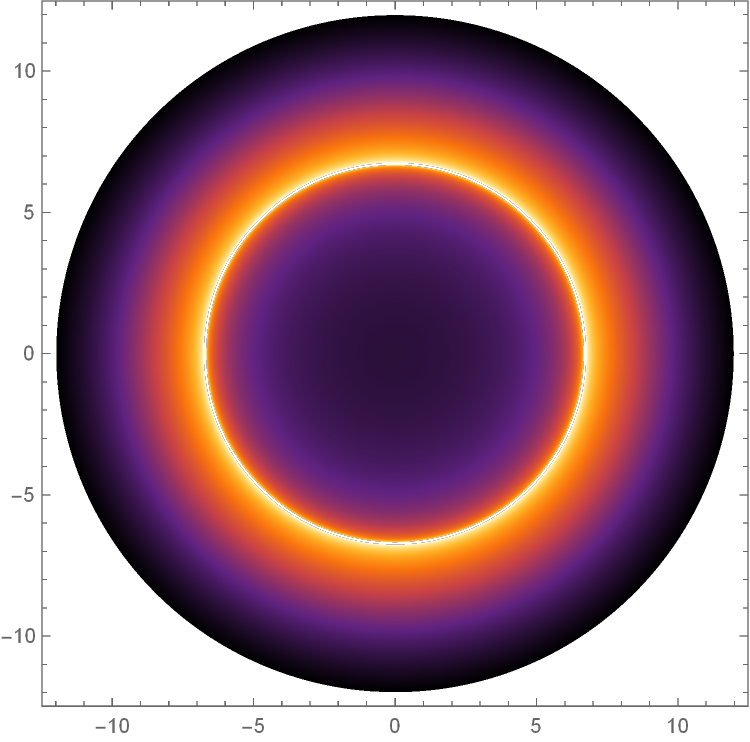}
\includegraphics[width=0.2in]{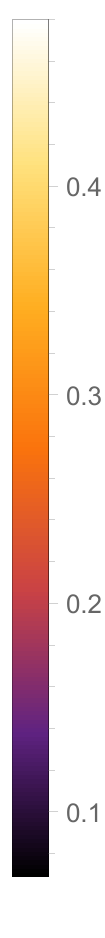}
\caption{Shadows and photon rings for static spherical accretion flow, $\omega = -0.5$ (top row) and $\omega = -0.7$ (bottom row) for different values of $g$, $a = 0.05$ and $M = 1$. The parameter $g$ from left to right is 0, 0.5 and 0.8, respectively.
}
\label{2-dimensional-2static}
\end{figure}

Shadows and photon rings of quintessence Hayward black holes for different values of $g$ with $\omega=-0.5$ and $\omega=-0.7$ are shown in Fig.~\ref{2-dimensional-2static}. It is clear that for a fixed value of $\omega$ with increase of $g$, the radius of shadows and photon rings of quintessence Hayward black holes decrease while their luminosity increases, so that the quintessence Schwarzschild black hole with $g=0$ has the lowest luminosity in comparison with quintessence Hayward black holes. This is due to the fact that the gravitational field of quintessence Schwarzschild black hole with $g=0$ is stronger which increases the light deflection and leads to more photons being trapped by the black hole and a lower luminosity of shadow and photon ring being observed compared to quintessence Hayward black holes. Also, comparing plots in the top row of the figure with those in the bottom row shows the effect of the state parameter of the quintessence matter $\omega$ on the black hole shadow and photon ring; for a fixed value of $g$ the luminosity of the photon ring increase with the absolute value of $\omega$.

\subsection{The infalling spherical accretion}
Since in our real universe the most of accretions are moving, in this section we consider the black hole shadows and photon rings in the context of infalling spherical accretion. Note that the observed intensity for a distant observer in the case of infalling spherical accretion is still expressed as equation (\ref{24}), but the redshift factor is different from static model and is related to the velocity of accretion flow
\begin{equation}
g = \frac{k_{\alpha}u_{\rm obs}^{\alpha}}{k_{\beta}u_{\rm em}^{\beta}},
\label{25}
\end{equation}
where $k^{\mu}\equiv {\dot{x}}^{\mu}$, $u_{\rm obs}^{\mu}=(1,0,0,0)$ and $u_{\rm em}^{\mu}$ are the photon four-velocity, distant observer four-velocity and accretion four-velocity, respectively. From equation (\ref{8}) we know that $k_t=1/b$ is a constant and $k_r$ can be obtained from condition $k_{\mu}k^{\mu}=0$. Therefore, we find
\begin{equation}
\frac{k_r}{k_t} = \pm\sqrt{\frac{1}{f(r)}\left(\frac{1}{f(r)}-\frac{b^2}{r^2}\right)},
\label{26}
\end{equation}
where the upper and lower sign denote to the case that the photons approach or away from the black hole, respectively. The four-velocity of the infalling accretion is $(u_{em}^{t}, u_{em}^{r}, u_{em}^{\theta}, u_{em}^{\varphi} )=(\frac{1}{f(r)}, -\sqrt{1-f(r)}, 0, 0)$. Using these equations, the redshift factor in equation (\ref{25}) can be obtained as
\begin{equation}
g = \frac{1}{u_{em}^t+\left(\frac{k_r}{k_t}\right)u_{em}^r}.
\label{28}
\end{equation}
Moreover, the proper distance is defined as
\begin{equation}
dl_p = k_{\alpha}u^{\alpha}_{em}ds = \frac{k_t}{g^3 \mid k_r\mid}dr,
\label{30}
\end{equation}
where $s$ is the affine parameter along the photon path. By assuming that the specific emissivity is monochromatic the specific intensity $I(\nu_{\rm obs})$ in the case of infalling spherical accretion thus can be written as
\begin{equation}
I = \int \frac{g^3}{r^2\sqrt{\frac{1}{f(r)}\left(\frac{1}{f(r)}-\frac{b^2}{r^2}\right)}} dr.
\label{31}
\end{equation}

\begin{figure}[H]
\centering
\includegraphics[width=3.0in]{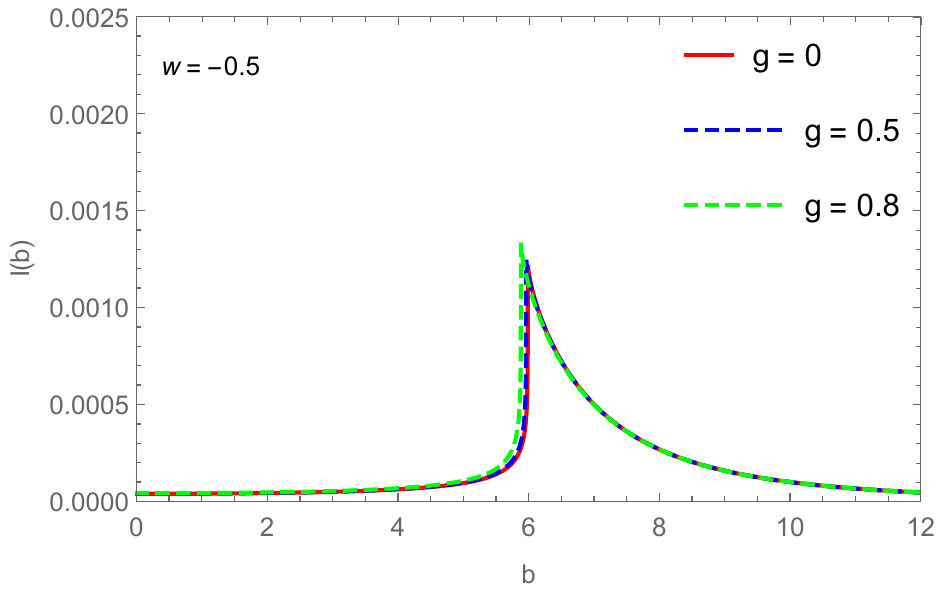}
\includegraphics[width=3.0in]{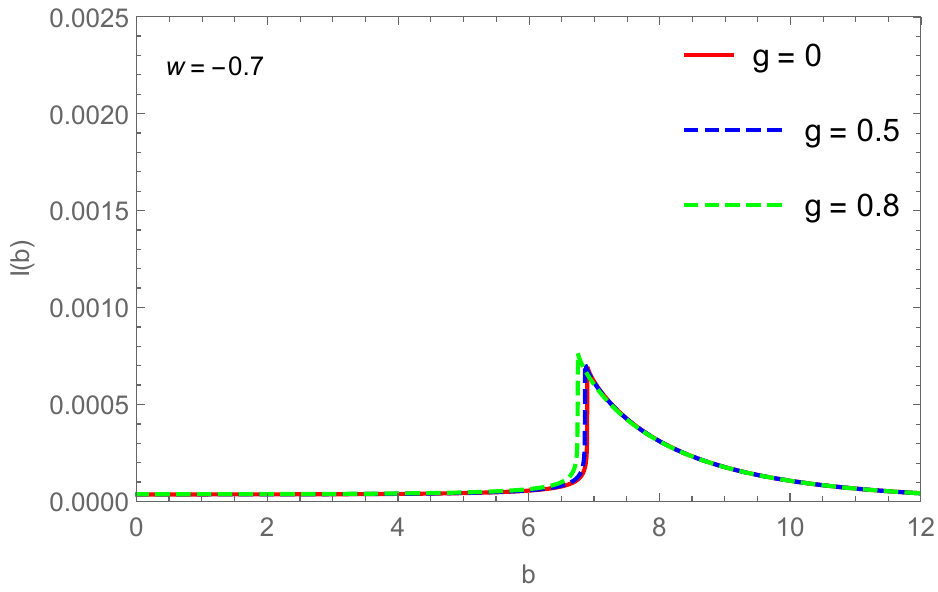}
\caption{ The observed intensity $I(\nu_{\rm obs})$ for infalling spherical accretion flow around a quintessence Hayward black hole $\omega = -0.5$ (left panel) and $\omega = -0.7$ (right panel) for different value of $g$, $a = 0.05$ and $M = 1$.}
\label{intensity-infalling}
\end{figure}

The observed intensity with respect to parameter $b$ with $\omega=-0.5$ (left panel) and $\omega=-0.7$ (right panel) is plotted in Fig.~\ref{intensity-infalling}. It is easy to see that the behavior of intensity is similar to the case of static spherical accretion flow and peak is at $b_{ph}$, but the maximum of intensity in this case is less than that of static case.

\begin{figure}[H]
\centering
\includegraphics[width=1.7in]{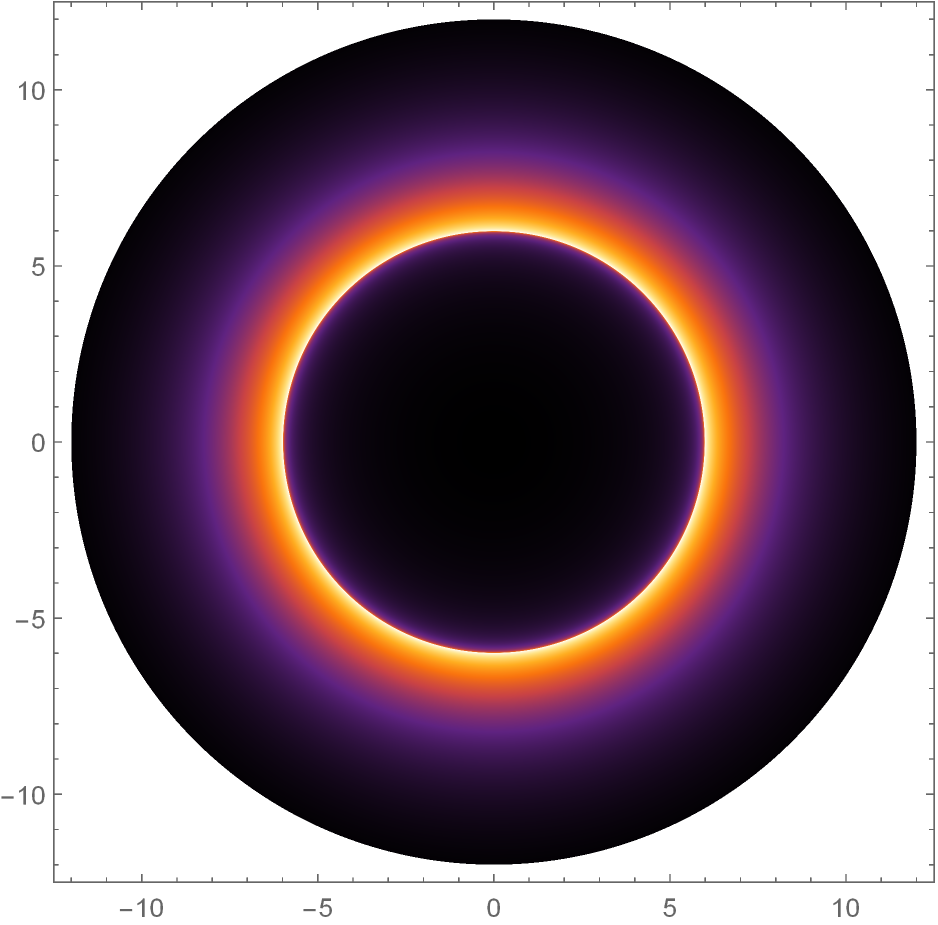}
\includegraphics[width=0.26in]{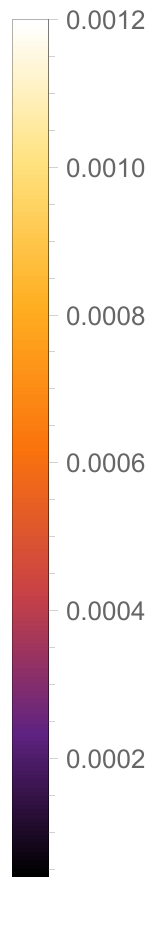}
\includegraphics[width=1.7in]{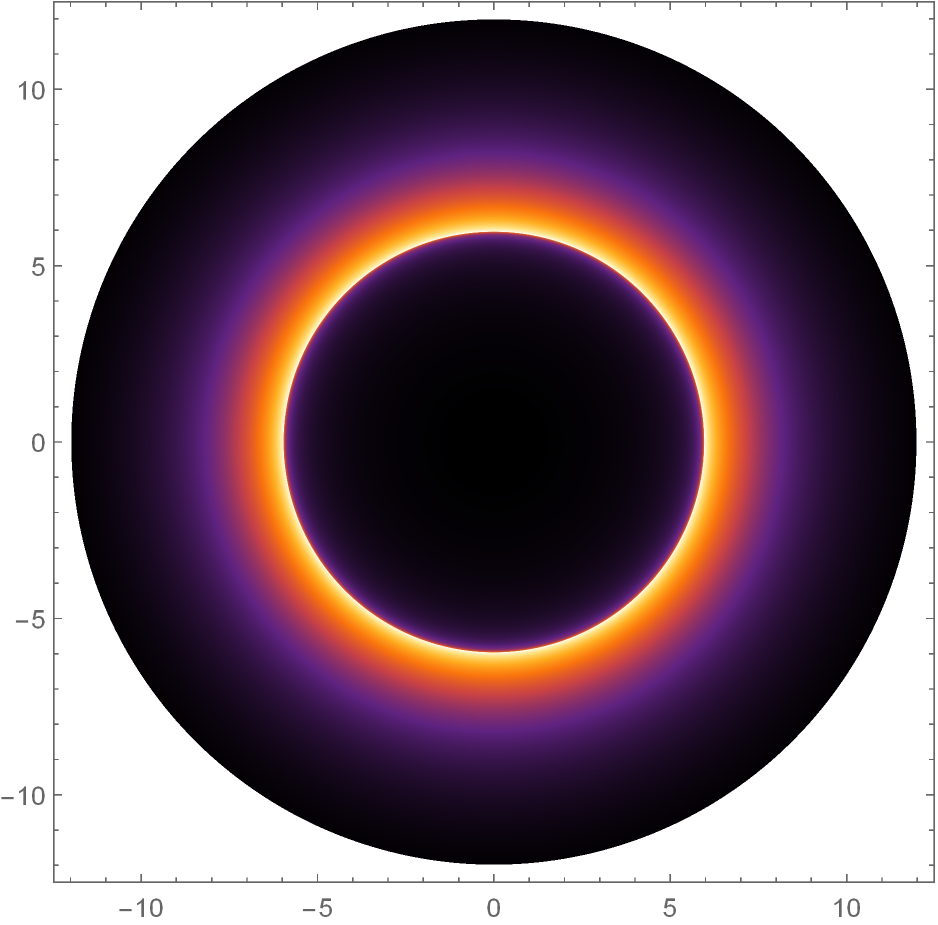}
\includegraphics[width=0.26in]{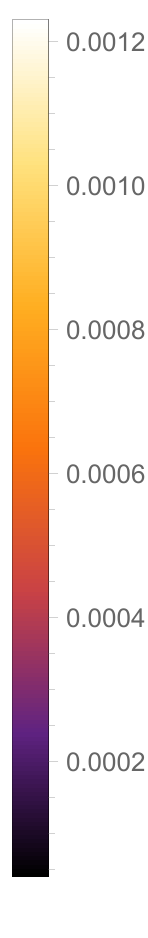}
\includegraphics[width=1.7in]{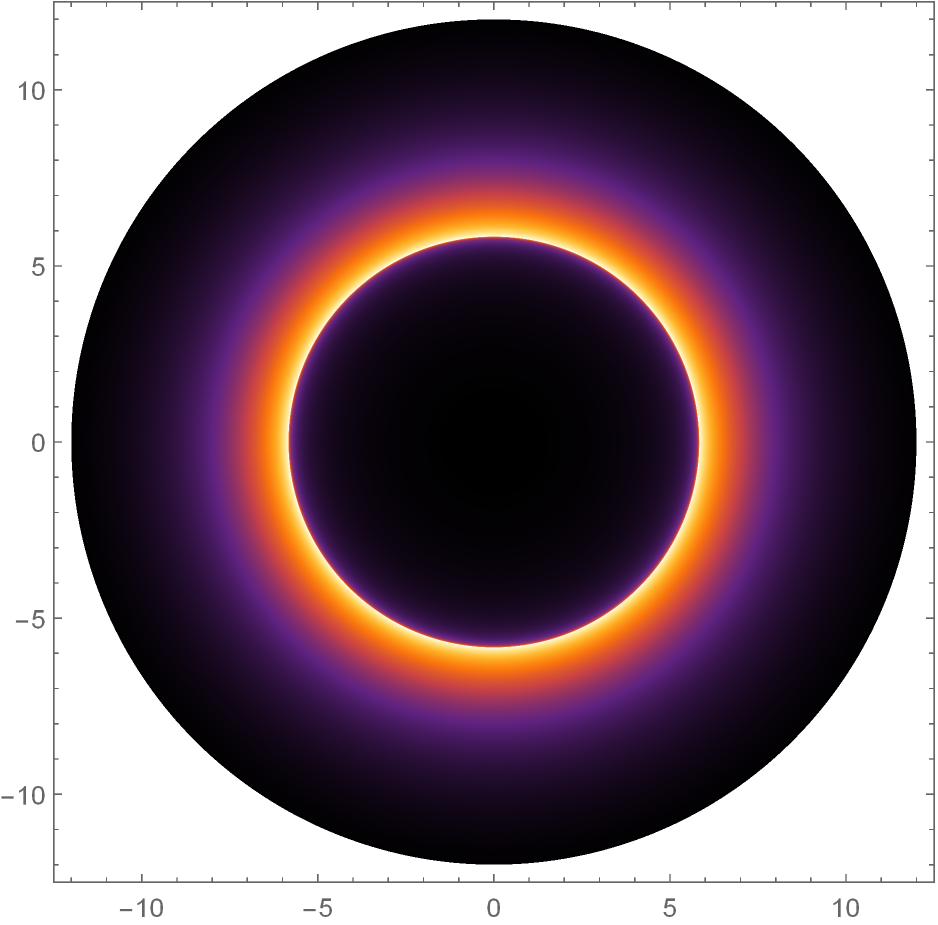}
\includegraphics[width=0.27in]{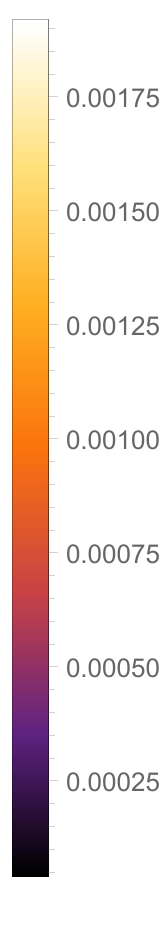}
\includegraphics[width=1.7in]{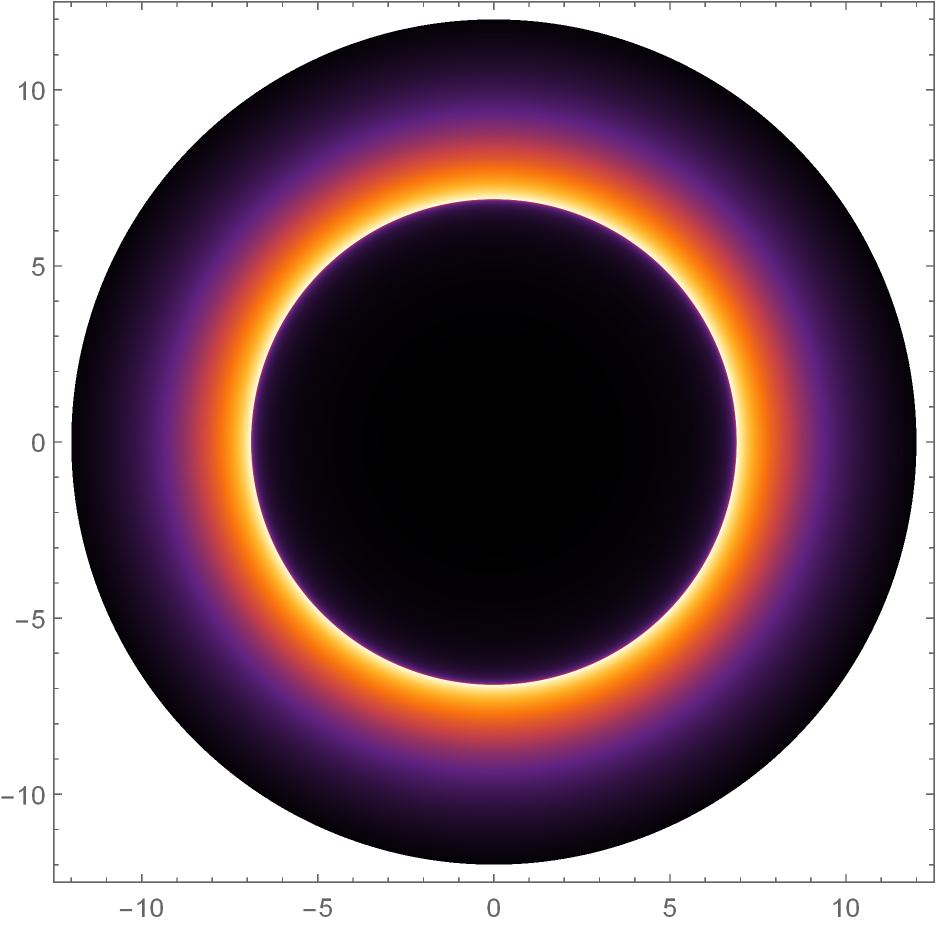}
\includegraphics[width=0.26in]{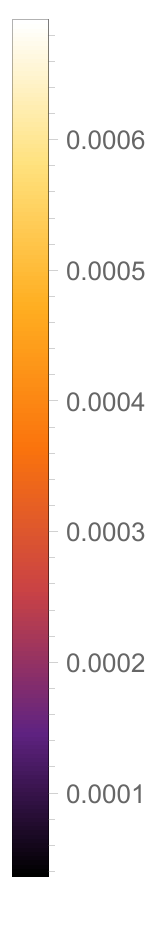}
\includegraphics[width=1.7in]{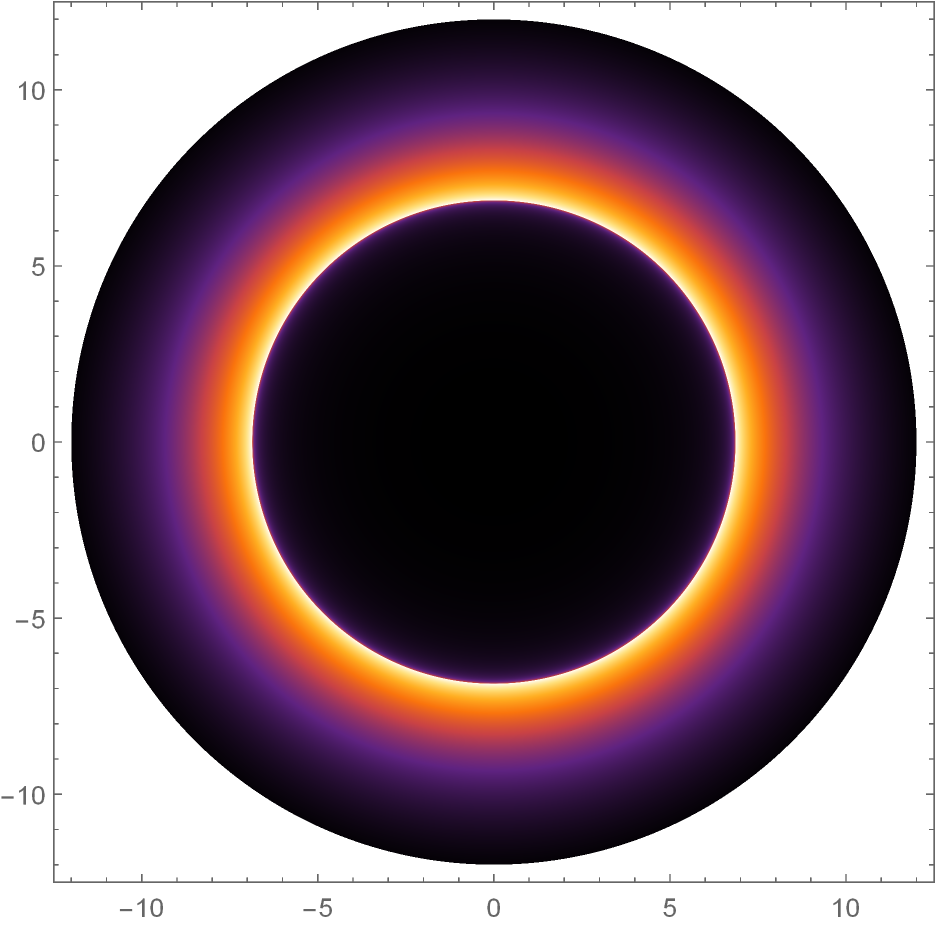}
\includegraphics[width=0.26in]{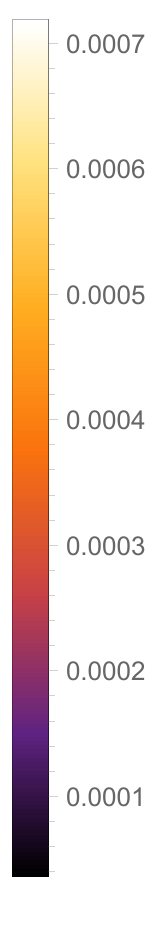}
\includegraphics[width=1.7in]{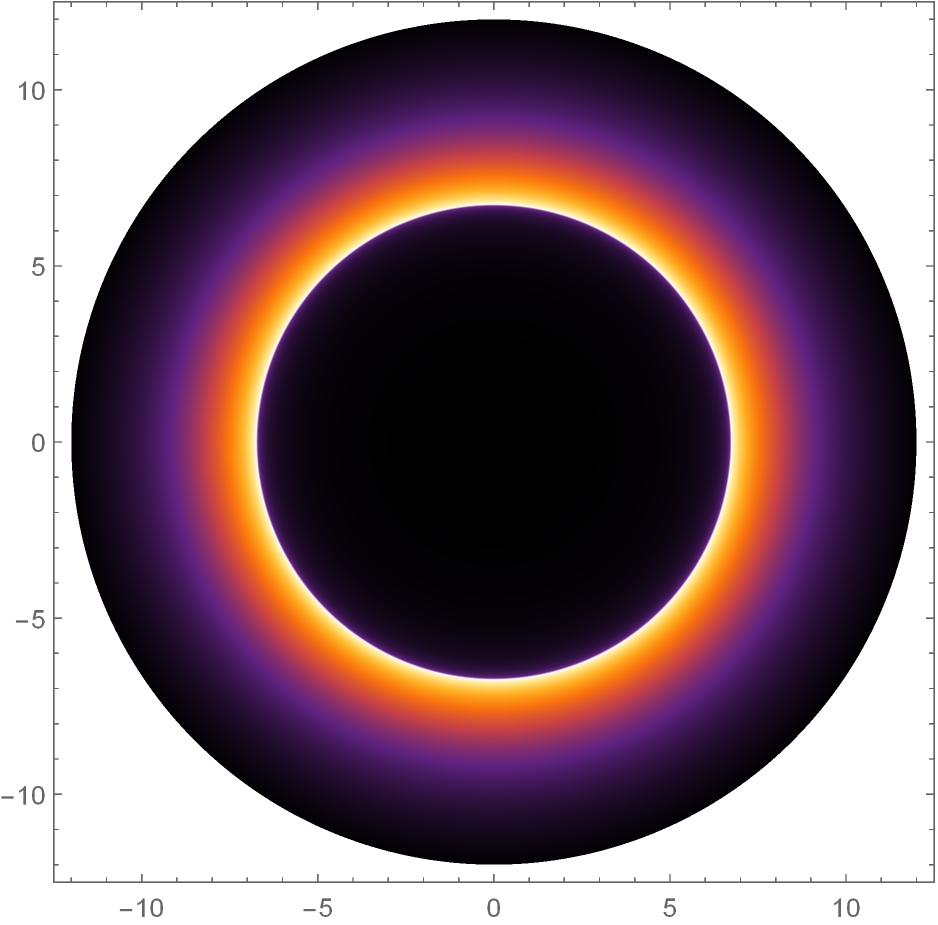}
\includegraphics[width=0.26in]{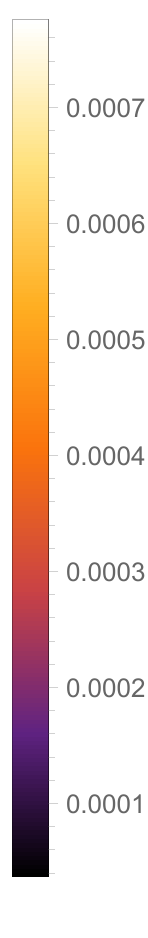}
\caption{ Black hole shadows and photon rings for infalling spherical accretion flow, $\omega = -0.5$ (top row) and $\omega = -0.7$ (bottom row) for different value of $g$, $a = 0.05$ and $M = 1$. The parameter $g$ from left to right is 0, 0.5 and 0.8, respectively}
\label{2-dimensional-2infalling}
\end{figure}

\begin{figure}[H]
\centering
\includegraphics[width=3.0in]{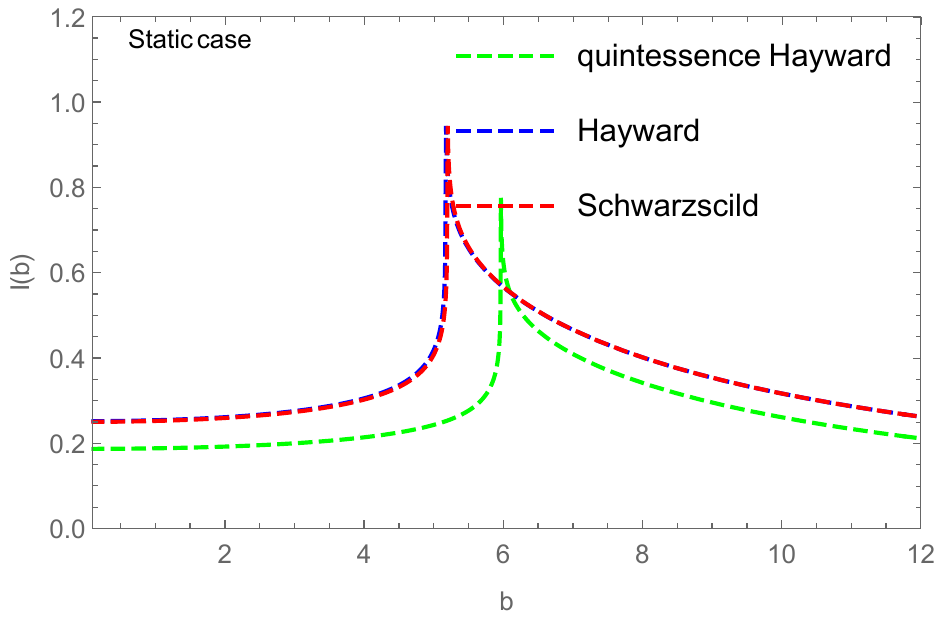}
\includegraphics[width=3.0in]{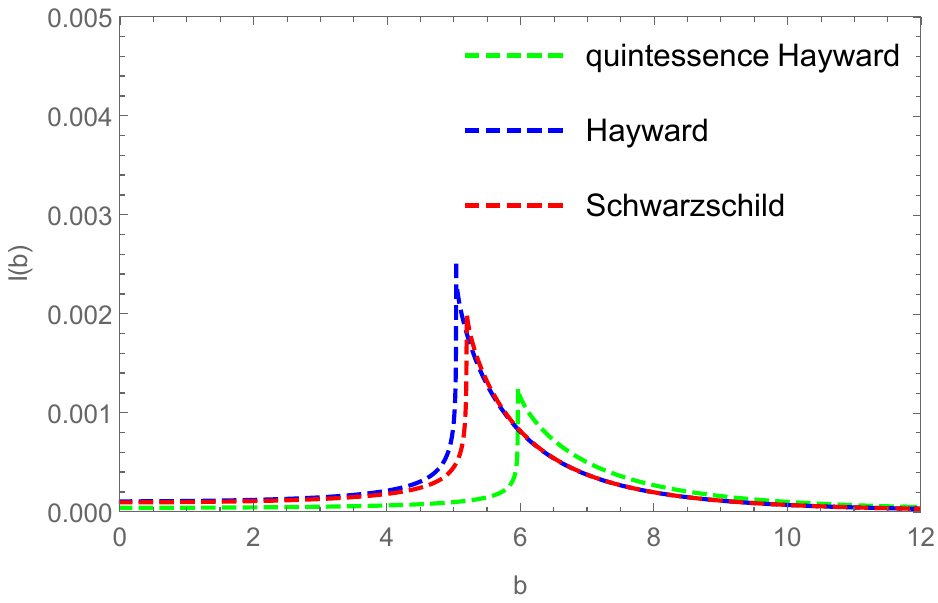}
\caption{ The observed intensity $I(\nu_{\rm obs})$ for static (left panel) and infalling (right panel) spherical accretion flow around a Schwarzschild, Hayward and quintessence Hayward black hole for $g = 0.9$, $a = 0.05$ and $M = 1$.}
\label{icompare-infalling}
\end{figure}

Image of shadow cast for $\omega=-0.5$ and $\omega=-0.7$ are shown in Fig.~\ref{2-dimensional-2infalling}. Clearly, in both cases at fixed $\omega$ with increasing parameter $g$, luminosity of shadows and photon rings increase. Moreover, by comparing plots in the top row of the figure with those in the bottom row we find that at a given $g$, for instance $g=0.8$, the radius of shadows and photon rings and their luminosities increase with the absolute value of state parameter. Note that the properties of the black hole shadows and photon ring not only depend on the space-time geometry but also on the accretion flow property. In the other words, the radius of shadows and photon rings do not change in the case of static and infalling spherical accretion, but we find that the specific intensity of static accretion is higher than that of infalling accretion for the same parameters $\omega$ and $g$, resulting in the central shadow region for the infalling accretion is darker than that for static accretion which is caused by the Doppler effect in this case. Finally, the observed intensity in the space-times of Schwarzschild black hole, Hayward and quintessence Hayward black holes has been compared in the left and right panels of Fig.~\ref{icompare-infalling} for static and infalling cases, respectively.

\section{Conclusions}
\label{1-Conclusions}
Shadows and observational appearance of quintessence Schwarzschild black hole surrounded by various profiles of accretion flow has been studied in \cite{Zeng1}. On the other hand the analysis of null geodesics for quintessence Hayward black hole (without accretion flow) is explored in \cite{Pedraza}. In this paper, we considered quintessence Hayward black hole solution \cite{Pedraza} surrounded by the static/infalling spherical accretion and similar to Ref. \cite{Zeng1} investigated the effect of accretion flow on the optical appearance of this black hole. First, we discussed the event horizon properties of quintessence Hayward black hole. We calculated the numerical values of the inner horizon $r_-$, event horizon $r_h$, cosmological horizon $r_c$, the radius of the photon sphere $r_{ph}$, and impact parameter of the photon sphere $b_{ph}$ for various values of $g$ and $\omega$, with the results summarized in Table~\ref{T2} . According to Table~\ref{T2} for a fixed $\omega$ with increasing $g$ the values of $r_{ph}$ and $b_{ph}$ decrease, while at fixed $g$ when we increase the absolute value of $\omega$, the values of $r_{ph}$ and $b_{ph}$ increases too. Photon trajectories in the vicinity of this black hole for different values of energy are plotted and showed the presence of quintessence matter increases the deflection of light rays, while the parameter $g$ oppositely effects the behavior of light deflection. We also constrained the amount of the free parameter of the quintessence Schwarzschild black hole with $\omega=-0.7$ compatible with the EHT observations of Sgr A* and M87* supermassive black holes.

Furthermore, we assumed that quintessence Hayward black hole surrounded by spherical accretion flow and considered two relativistic models: the static and infalling spherical accretion flow. Then, in Fig. \ref{intensity-static} and Fig. \ref{intensity-infalling} we plotted the profile of observed intensity as a function of impact parameter for $\omega=-0.5$ and $\omega=-0.7$ with different values of $g$, in the model with gas at rest and, in radially infalling gas accretion scenario, respectively. Shadows and photon rings for $\omega=-0.5$ and $\omega=-0.7$ with different values of $g$ are also shown in Fig. \ref{2-dimensional-2static} and Fig. \ref{2-dimensional-2infalling}. The size of shadows and photon rings decrease and their intensity and luminosity increase for a fixed $\omega$ with increasing $g$, while the effect of the absolute value of $\omega$ is opposite. For a fixed $\omega$ with decreasing $g$, the strength of the gravitational field around the black hole increases and so the impact parameter increases and more photons capture by the black hole (tiny fraction of photons escape from black hole), and so a lower luminosities of shadows and photon rings being observed. Also, the results of quintessence Hayward black holes shows that the effect of quintessence background on the size of the black hole shadow, photon rings and their luminosities is more significant in comparison with that for Hayward black holes while, the comparison between quintessence Schwarzschild and quintessence Hayward black holes shows that the regularity faintly affects the shadow size, photon rings and their luminosities. Finally, it is worth noting that the realistic black holes rotate so we will extend this static solution to kerr-like solution in future.

\section*{Data Availability}
Data sharing not applicable to this article as no datasets were generated or analysed during the current study.

\end{document}